\documentclass[11pt,twoside]{article}

\usepackage{asp2004}
\usepackage{epsf}
\usepackage{lscape}

\begin{document}
\title{Variation of the cluster luminosity function across the disk of M51}    
\author{  M.R. Haas,
          M. Gieles,
 	  R.A. Scheepmaker,
	  S.S. Larsen,
	  H.J.G.L.M. Lamers,
	  N. Bastian
          }   
\affil{Astronomical Institute, Utrecht University, Princetonplein 5, NL-3584 CC Utrecht, The Netherlands \\
       }    

\begin{abstract} 
We study the luminosity function (LF) of the star clusters in M51. Comparing the observed LF with the LF resulting from artificial cluster populations suggests that there exists an upper mass limit for clusters and that this limit and/or the cluster disruption varies with galactocentric distance.
\end{abstract}

\section{Introduction}
In cluster populations with a power law cluster initial mass function (CIMF) the highest appearing mass in a sample is determined by the number of clusters (besides statistical fluctuations). This is no longer true if there exists a physical upper mass limit for clusters.

Information about the CIMF can be derived from the present day luminosity function (LF). Although there is not a one-to-one relation between the CIMF and the LF, modeling the LF with an artificial cluster population with varying CIMF parameters can aid the interpretation of the LF. According to the models of \cite{gieles06}, a bend in the LF (i.e. a double power law distribution function) may be the signature of a CIMF, truncated at the high mass end. The (filter dependent) location of the bend is intimately connected to the value of the upper mass limit (brighter bends correspond to higher possible cluster masses).

Bends in the LF are observed in NGC 6946, M51 \cite{gieles06, gieles06a} and NGC 4038/4039 \cite{whitmore99}. Here we again examine the cluster population of M51, dividing it in subsamples at different galactocentric radii. By comparing to the artificial populations we draw conclusions on the CIMF and cluster disruption parameters across the disk. 

\section{The cluster population of M51}
We use the Hubble Heritage ACS mosaic of M51 in the passbands \textit{F435W}, \textit{F555W} and \textit{F814W}, covering the complete system of M51 and its companion \cite{mutchler05}. Point sources are extracted and qualified as a cluster if (1) the source is detected in all three broadband filters, (2) the radius of the source was fit better with an extended cluster profile than with a delta function (using the \textit{ISHAPE} algorithm \cite{larsen99}) (3) the cluster has a radius of at least 0.5 pc (the resolution of the \textit{ISHAPE} routine), and (4) the nearest neighbouring source is at least 5 pixels away.

After performing aperture photometry on all sources, the LF can be 
constructed for the population brighter than the 95\% completeness limit  (-6.3 abs mag \textit{F435W} and \textit{F555W} and -6.0 abs mag for \textit{F814W}) and for three galactocentric distance intervals. Results can be seen in Table~\ref{tab:fits}.

\begin{table*}
\caption{Fit results of the complete sample in all three pass bands. (1) is the passband, (2) the number of clusters within the fit range. (3), (4) and (5) contain both slopes and the location of the bend of the double power law fit respectively. The second part contains for the B band the galactocentric distance dependence. Similar results are obtained in the other filters.}
\label{tab:fits}      
\centering          
\begin{tabular}{l l l l l}     
\hline\hline
Passband & N  &  $\alpha_1$ & $\alpha_2$ & $M_{\textrm{\tiny{bend}}}$ \\
\hline                    
\textit{F435W} & 3891 & 1.96 $\pm$ 0.04 & 2.52 $\pm$ 0.08 & -8.33 $\pm$ 0.15\\
\textit{F555W} & 4750 & 1.99 $\pm$ 0.04 & 2.56 $\pm$ 0.07 & -8.38 $\pm$ 0.13\\
\textit{F814W} & 8041 & 2.08 $\pm$ 0.02 & 2.54 $\pm$ 0.08 & -8.90 $\pm$ 0.16\\
\hline                  
\hline
$F435W$: & N &  $\alpha_1$ & $\alpha_2$ & $M_{\textrm{\tiny{bend}}}$ \\
\hline
0 $< d <$ 3 kpc & 1267 & 1.67 $\pm$ 0.06 & 2.60 $\pm$ 0.17 & -8.76 $\pm$ 0.17\\
3 $< d <$ 5.5 kpc & 1415 & 2.08 $\pm$ 0.05 & 2.71 $\pm$ 0.22 & -8.42 $\pm$ 0.22\\
5.5 $< d <$ 8.5 kpc & 1209 & 2.17 $\pm$ 0.03 & 2.55 $\pm$ 0.31 & -7.99 $\pm$ 0.31\\
\hline
\end{tabular}
\end{table*}

\section{Conclusions on upper mass limits and disruption parameters}
Although more detailed modeling is necessary for quantitative results, the conclusions following the models of \cite{gieles06} for the galactocentric dependence of the upper mass limit and disruption parameters are:
\begin{enumerate}
\item Subsets closer to the center have their bends at brighter magnitudes, suggesting higher possible cluster masses.
\item Subsets closer to the center have shallower faint-end slopes, indicating faster disruption.
\end{enumerate} 
The two effects are not to be seen separately though. Faster disruption alters the location of the bend in the same direction as a higher upper mass limit. In a follow up study the statistical significance of the results will be further investigated and a more quantitative analysis will be given \cite{haas06}.



\begin{thebibliography}{}


\bibitem[Gieles et al. 2006a]{gieles06} Gieles, M., Larsen, S.~S., Anders, P., Bastian, N., \& Stein, I.~T.\ 2006, \aap, 450, 129
\bibitem[Gieles et al. 2006b]{gieles06a} Gieles, M., Larsen, S.~S., Scheepmaker, R.~A., Bastian, N., Haas, M.~R., \& Lamers, H.~J.~G.~L.~M.\ 2006, \aap, 446, L9
\bibitem[Haas et al. 2006]{haas06} Haas, M.R.,  Gieles, M., Scheepmaker, R.A., Larsen, S.S., Lamers, H.J.G.L.M., \& Bastian, N.\ 2006, \aap, in prep.
\bibitem[Larsen 1999]{larsen99} Larsen, S.~S.\ 1999, \aaps, 139, 393
\bibitem[Mutchler et al. 2005]{mutchler05} {Mutchler}, M., {Beckwith}, S.~V.~W., {Bond}, H., et al.,\ 2005 American Astronomical Society Meeting Abstracts, 206 
\bibitem[Whitmore et al. 1999]{whitmore99} Whitmore, B.~C., Zhang, Q., Leitherer, C., Fall, S.~M., Schweizer, F., \& Miller, B.~W.\ 1999, \aj, 118, 1551


\end{thebibliography}
\end{document}